\documentclass[a4paper,12pt]{article}
\usepackage{graphicx}
\rm
\begin{document}
\title{\bf\large LOW MASS EXOTIC BARYONS: MYTH OR REALITY ?}
\author{B. Tatischeff\\
   \\
Institut de Physique Nucl\'eaire, CNRS/IN2P3, F-91406
Orsay Cedex, France\\
E-mail: tati@ipno.in2p3.fr}
\date{ }
\maketitle
\begin{center}
{\bf\large Abstract}
\end{center}
\hspace*{4.mm}Using mainly $\vec{\mbox{p}}$~p $\rightarrow$ p ${\pi^+}X$ and
$\vec{\mbox{p}}$~p~$\rightarrow$~p$_{f}$~p$_{s}$~X reactions, narrow
baryonic structures
were observed in the mass range 950$\le$~M~$\le$~1800 MeV. 
\section{Introduction}
\hspace*{4.mm}Narrow structures in hadrons have been observed since two
decades.
Despite this long time these results are still in debate. Two reasons
explain, if not justify, such situation. First the observed structures are
often small, and some observations were made using bubble chamber slides,
therefore
with small statistics. The second reason is that their genuine existence 
constitute a disturbing element inside the present theories.\\
\hspace*{4.mm}Narrow structures in baryons were first observed in the
missing mass M$_{X}$ of the p~p~$\rightarrow$ p~$\pi^{+}$~X
reaction, in the
mass range 1.0$\le$~M~$\le$~1.1 GeV \cite{bor1}. The two lowest 
structure masses are small, preventing them from disintegration
through pion emission. When all three structures inside this mass range at
M=1004 MeV, 1044 MeV, and 1094 MeV are observed at nearly all angles and
energies, narrow structures are also observed in other
mass ranges, but they are more weakly excited, and therefore extracted with
less confidence.\\
\hspace*{4.mm}The experiment was performed at the SPES3 beam
line using the
Saturne polarized proton beam ($T_{p}$=1.52 GeV, 1.805 GeV,
and 2.1 GeV). Narrow
baryons were observed in the missing mass M$_{X}$ and in the invariant
masses
M$_{p\pi^+}$ and M$_{X\pi^+}$ of the p~p~$\rightarrow$~p~${\pi^+}X$
reaction, and in the invariant masses M$_{Xp_{f}}$ and M$_{Xp_{s}}$ of the
p~p~$\rightarrow$~p$_{f}$~p$_{s}$~X reaction. Here p$_{f}$ (p$_{s}$) means
 fast (slow) proton.
Some structures were also
observed at nearby masses in the missing mass of the
d~p~$\rightarrow$~p~p~X reaction.
Fig. 1 shows the mass ranges studied in the Saturne (SPES3) experiment. The
three vertical strips for each variable correspond to the three incident
energies, increasing from left
to right. The horizontal lines show the masses where a narrow structure was
observed.\\ 

\begin{center}
\begin{figure}[!h]
\vspace*{3.mm}
\scalebox{.9}[.6]{
\includegraphics[bb=10 30 500 520,clip,scale=1.]{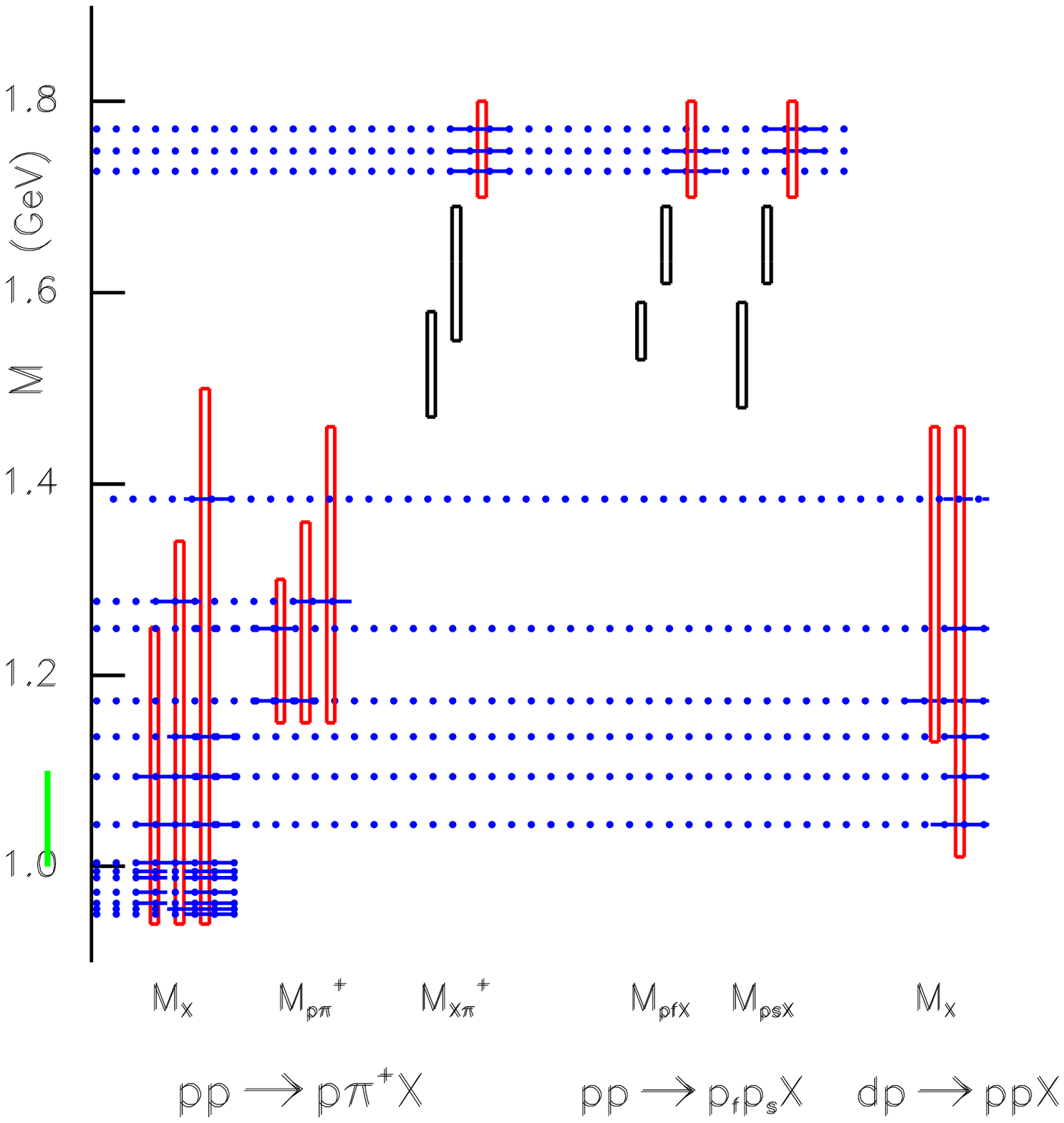}}
\vspace*{-3.mm}
\caption[Fig. 1]{Experimental ranges for the study of narrow baryonic masses
(in GeV).}
\end{figure}
\end{center}

\vspace*{.5cm}
\begin{center}
\begin{figure}[!h]
\scalebox{.9}[.6]{
\includegraphics[bb=40 340 580 833,clip,scale=1.]{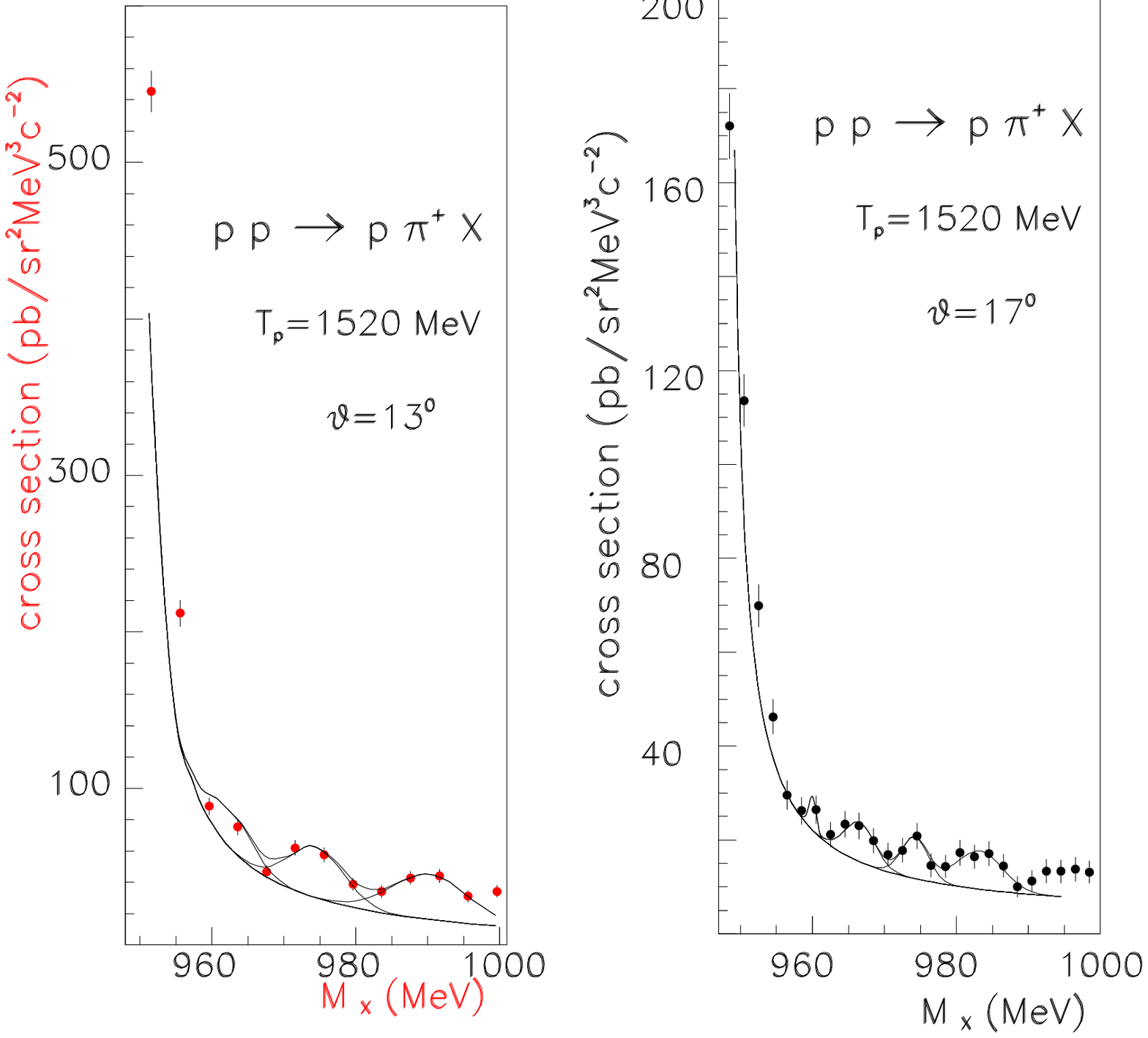}}
\vspace*{-3.mm}
\caption[Fig. 2]{Selection of two cross sections showing an oscillatory
pattern in the missing mass.}
\end{figure}
\end{center}

\vspace*{-1.7cm}
\section{Narrow baryons in the mass range 950$\le$~M~$\le$995 MeV}
\hspace*{4.mm}When the missing mass spectra display smooth patterns up to
M=950 MeV, they display oscillatory patterns for larger masses. An example
of such oscillations is shown in Fig. 2.\\

\hspace*{4.mm} A systematic control was undertaken
to make sure that these structures were not produced by the lost event
corrections or by the acceptance renormalizations. Fig. 3 shows that
the analyzing powers display - like the cross sections -  a non smooth behavior.
Here only small angle data are shown, since the error bars increase quickly
for angles $\ge~9^{0}$.\\

\begin{center}
\begin{figure}[!hb]
\scalebox{.9}[.6]{
\includegraphics[bb=13 13 513 513,clip,scale=1.]{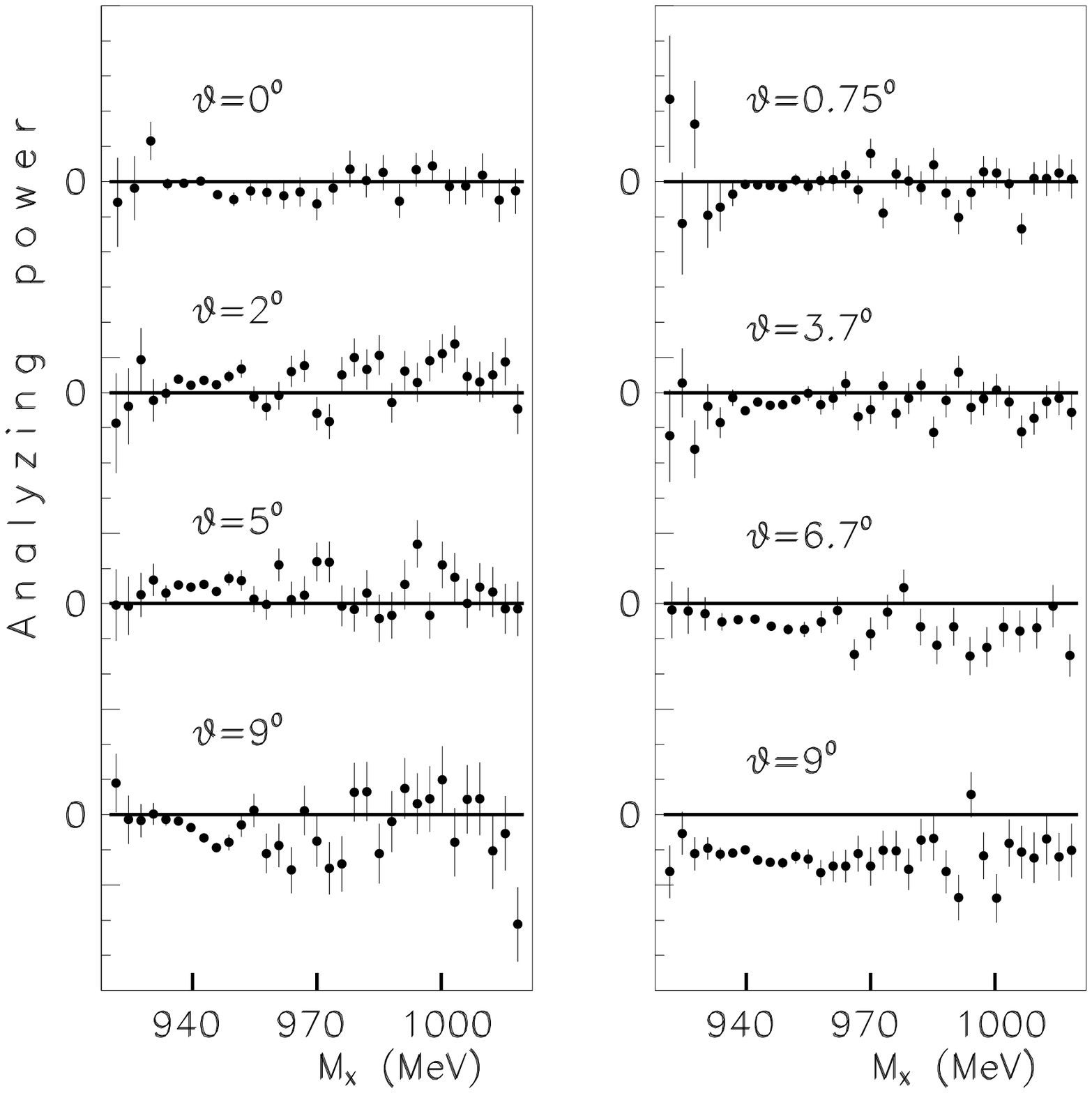}}
\caption[Fig. 3]{Selection of small angle analyzing powers of the missing
mass M$_{X}$ from the  $\vec{\mbox{p}}$~p~$\rightarrow$~p~${\pi^+}X$
reaction. The two
columns correspond from left to right to T$_{p}$=1520~MeV and
T$_{p}$=1805~MeV.}
\end{figure}
\end{center}
\hspace*{4.mm}For the observed oscillatory shapes in
cross sections, the background determination is
clearly somewhat ambiguous. Several different choices can be used for
background. The first
one consists to draw an averaged background inside the data. Then all structures
will be strongly reduced, if not disappear. However we consider as quit unlikely
the situation where nearly all spectra will exhibit an oscillatory pattern
and where all these oscillations will be accidental. The second choice consists
to consider these variations as physical, and to give them the same width as the
experimental neutron missing mass widths which increased slowly with the
spectrometer angle. Then the
structures will be extracted with a higher number of
standard deviation (S.D.). We choose the third intermediate choice and
draw the background using the low data points as in Fig. 2.\\
\hspace*{4.mm}Although 54 peaks were extracted from all cross
section spectra, only 18 were kept which had a
valuable statistical significance, since they were extracted with
S.D.$\ge$~3. The corresponding masses are displayed
in Fig. 4. When a mass ($\pm$~1~MeV) is obtained at least twice, it is
considered as being a candidate for a new structure and an horizontal dashed
range is drawn in Fig. 4. The figure shows also the two narrow masses (at 966 MeV and
986 MeV), extracted by L.V. Filkov {\it et al.} from the
p~d~$\rightarrow$~p~p~X$_{1}$ reaction \cite{fil}.

\begin{center}
\begin{figure}[!ht]
\scalebox{.9}[.6]{
\includegraphics[bb=27 18 517 531,clip,scale=1.]{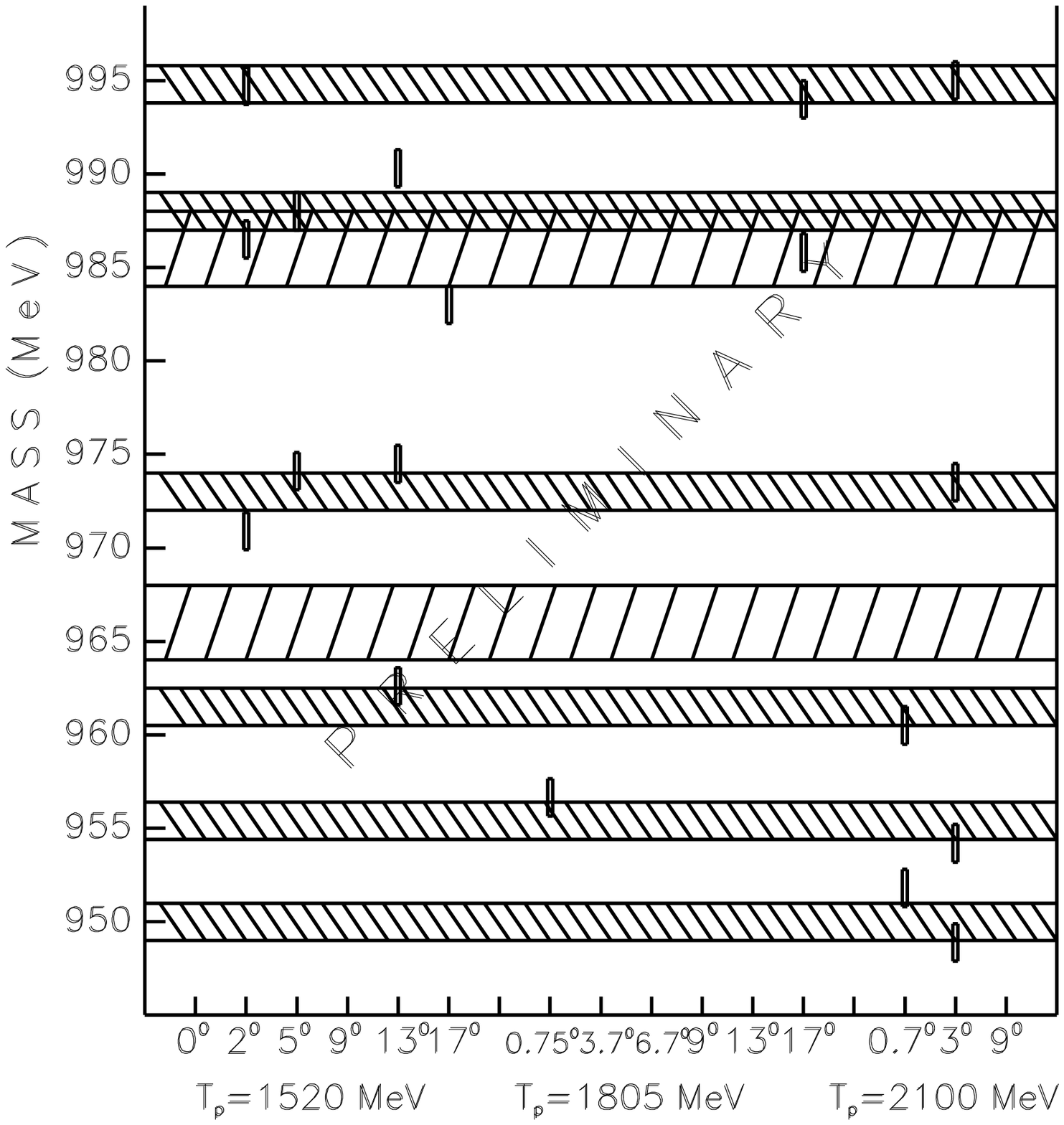}}
\caption[Fig. 4]{Low narrow exotic baryonic masses.}
\end{figure}
\end{center}

\hspace*{1.mm}We tentatively associate these structures to exotic baryons
formed by two coloured quark clusters, and we use phenomenologically a mass
formula derived 23 years ago \cite{mul}:
\begin{equation}
\hspace*{-1.cm}
M = M_0+M_1[i_1(i_1+1)+i_2(i_2+1)+(1/3)s_1(s_1+1)+
(1/3)s_2(s_2+1)]
\end{equation}
where s$_{1}$(s$_{2}$) and i$_{1}$(i$_{2}$) are the first (second)
cluster spin and isospin values. The two parameters, M$_{0}$=924
MeV,
and M$_{1}$=15.6 MeV are adjusted in order to describe the neutron mass and
the first exotic mass at 950 MeV. The other masses, possible spins and
isospins obtained with the use of this formula, are shown in Fig. 5.
We observe the good
correspondance between the experimental and the calculated masses.\\

\begin{center}
\begin{figure}[!ht]
\hspace*{2.cm}
\scalebox{.9}[.6]{
\includegraphics[bb=3 32 420 490,clip,scale=0.88]{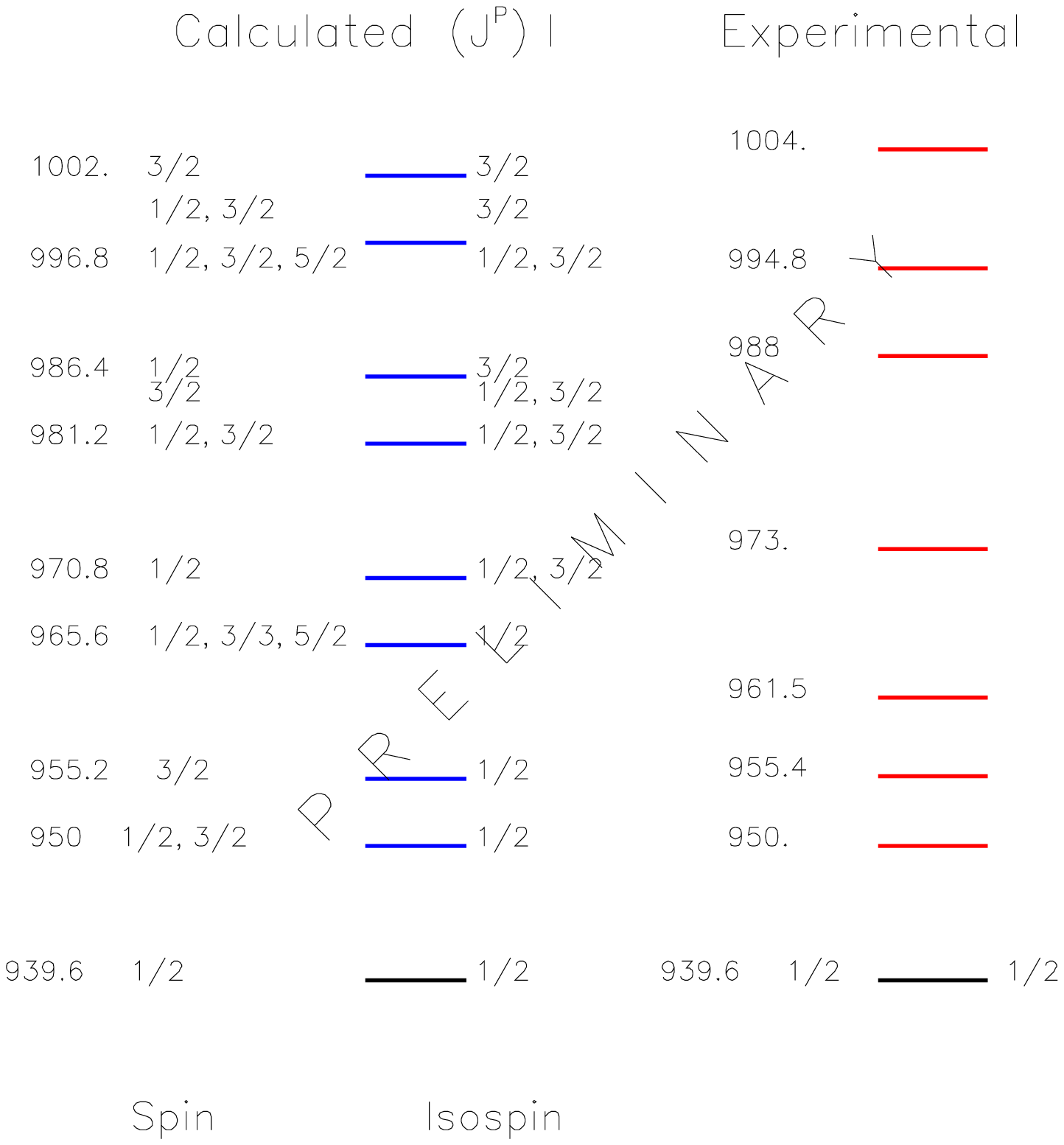}}
\caption[Fig. 5]{Experimental and calculated exotic baryonic masses in the
range 948~$\le$~M~$\le$~995~MeV.}
\end{figure}
\end{center}

\begin{center}
\begin{figure}[!hb]
\hspace*{1.5cm}
\scalebox{.9}[.6]{
\includegraphics[bb=54 364 520 780,clip,scale=0.88]{/home/tati/bar191.eps}}
\caption[Fig. 6]{Selection of four spectra illustrating a narrow structure
close to M$_{X}$=1136 MeV.}
\end{figure}
\end{center}

\vspace*{-1.5cm}
\section{Narrow baryons in the mass range 1100$\le$~M~$\le$1400 MeV}
\hspace*{4.mm}Some results corresponding to this mass range were already
reported \cite{bor2}. Narrow peaks were extracted
from missing mass or invariant mass spectra.
Only those structures observed several times at the same mass $\pm$~5~MeV
were taken into account. A selection of some spectra 
showing 4 peaks close to 1136 MeV is presented in Fig. 6.\\
\hspace*{4.mm}Several
structures were observed in this mass range, extracted from the experiments
quoted previously, and also from the invariant mass M$_{p\pi}$ of
the $\gamma$~n~$\rightarrow~\pi^{0}~\pi^{-}$~p reaction studied at MAMI
\cite{zab}, and from the center of mass energy of the
$\gamma$~p~$\rightarrow~\pi^{+}$~n reaction studied at Bonn \cite{dan}. The
results are all listed in Fig. 7 which is explained~in~table~1.\\
\hspace*{4.mm}The same mass formula as the one discussed before, was used
in this mass range. The two parameters M$_{0}$=838.2~MeV and
M$_{1}$=100.3~MeV
were adjusted in order to describe the mass (and possible spin and isospin)
of the neutron and of the Roper N(1440 MeV)
 resonance. Fig. 8 shows the good
 agreement between the experimental and the calculated masses. The last
 column in Fig. 8 shows the masses calculated by N. Konno, H. Noya and H.
 Nakamura \cite{kon} inside their diquark cluster model.\\
\begin{center}
\begin{figure}[!ht]

\vspace*{-.8cm}
\hspace*{2.cm}
\scalebox{1.}[.8]{
\includegraphics[bb=10. 10. 530. 530.,clip,scale=0.6]{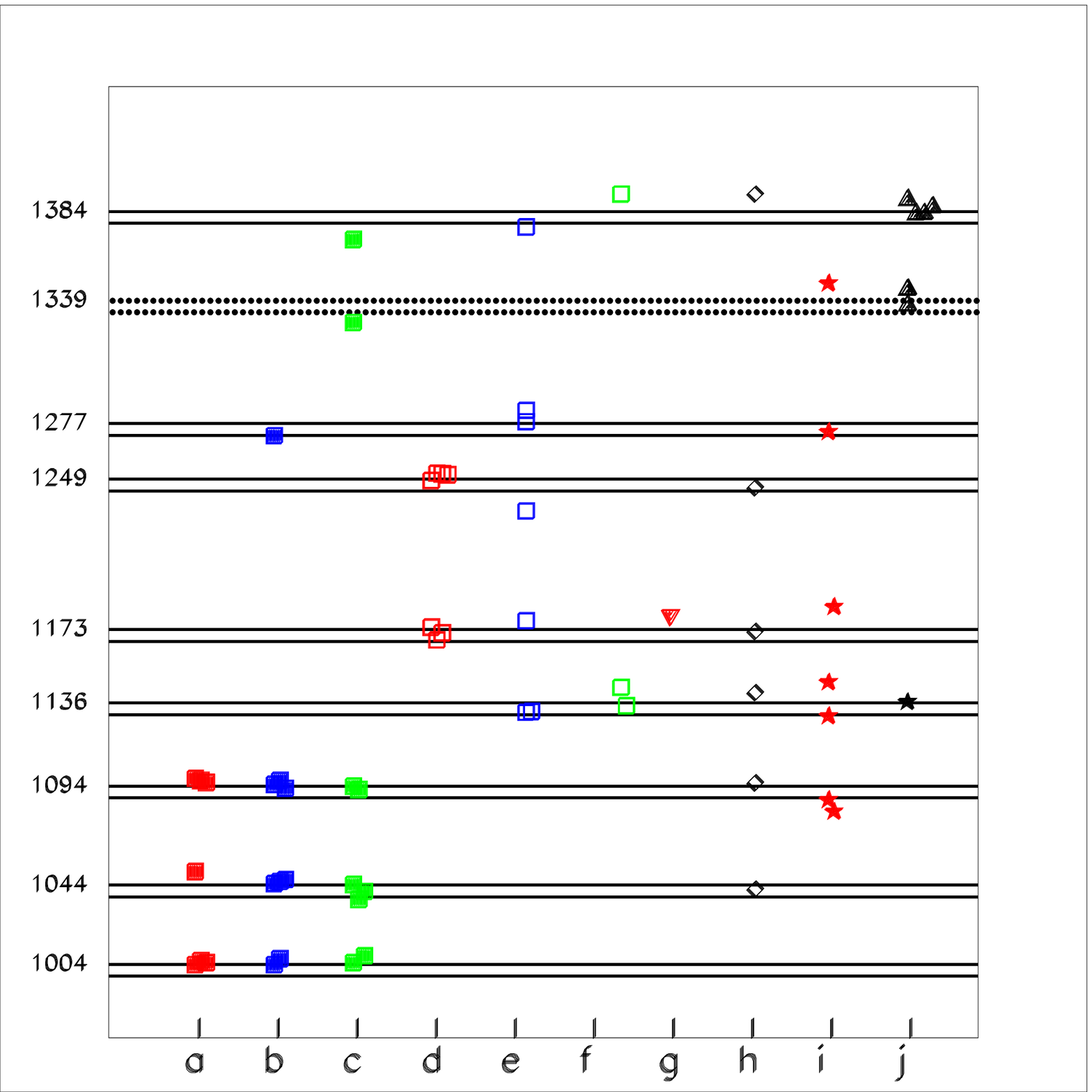}}
\caption[Fig. 7]{Narrow structure baryonic masses observed in cross sections
from different reactions in the range 1100~$\le$~M~$\le$~1400 MeV.}
\end{figure}
\end{center}

\vspace*{-1.5cm}
\section{Narrow baryons in the mass range 1500$\le$~M~$\le$1700 MeV}
\hspace*{4.mm}Starting from this mass range, we expect to observe a complex
situation, due to a superposition of weakly excited structures as those
discussed previously with interferences from many broad overlapping 
PDG resonances having the same
quantum numbers. Many such resonances overlap, as N$^{*}$, J$^{P}$=1/2$^{-}$,
1/2$^{+}$, 3/2$^{+}$, or $\Delta$, J$^{P}$=3/2$^{+}$, 1/2$^{-}$,3/2$^{-}$,
and 1/2$^{+}$ in the mass range studied here. We observe indeed structures 
slipping smoothly in mass for increasing spectrometer angles, superimposed
to structures stable in mass. The data analysis in this mass range is not
completed.\\
\newpage
\begin{center}
\hspace*{5.cm}
\begin{table}[!ht]

\vspace*{-1.cm}
\scalebox{1.1}[1.]{
\includegraphics[bb=40 506 523 794,clip,scale=0.88]{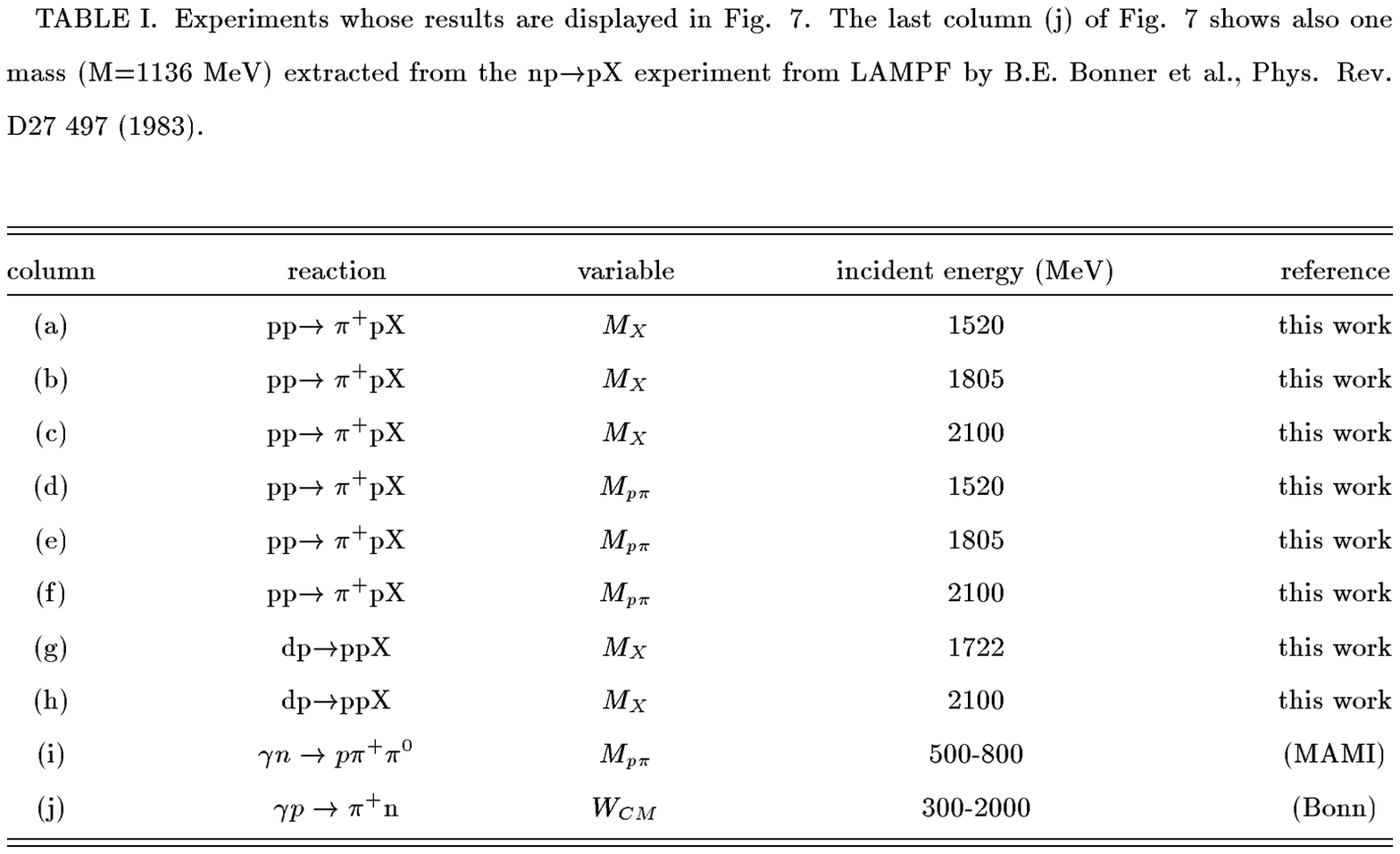}}
\end{table}
\end{center}

\vspace*{-2.mm}
\section{Narrow baryons in the mass range 1720$\le$~M~$\le$1800 MeV}
\hspace*{4.mm}Surprisingly, the superposition of smoothly moving structures
with structures stable in mass (those we are looking for here) seems not
impede us in this mass range. Three structures at M=1727 MeV, 1748 MeV, and
1771 MeV were observed at different angles, in different variables from both
reactions using our highest incident proton energy: T$_{p}$=2.1 GeV.
These masses were observed in ten different experimental conditions, always 
- but in one case - with SD$\ge$3. Fig. 9 shows some illustrations of
these results. The insert (d) shows an analyzing power data where narrow
structures are present at the same masses where they appear in cross
sections, namely at 1727 MeV and 1748 MeV.\\
\section{Conclusion}
\hspace*{4.mm}Narrow, low mass, non strange structures were observed in
baryonic missing masses and in baryonic
invariant masses studied with help of different reactions. These weakly
excited narrow structures
were observed in the whole baryonic mass range investigated: 
940$\le$~M~$\le$~1800~MeV.\\
\hspace*{4.mm}Many checks were undertaken
to make sure that these structures were not produced by experimental
artifacts \cite{bor3}. Their small widths and the stability of the extracted masses,
regardless of the experiment, of the variable under study, of the incident
energy, and of the
reaction angle (see Fig. 1), allowed us to conclude that
these structures are genuine and not produced by dynamical rescatterings.
We concluded that these peaks correspond to new exotic baryons since
there is no room for them within the many theoretical constituent quark
models \cite{cap}.\\

\begin{center}
\begin{figure}[!ht]
\hspace*{7.mm}
\scalebox{0.8}[.8]{
\includegraphics[bb=0 50 534 538,clip,scale=0.8]{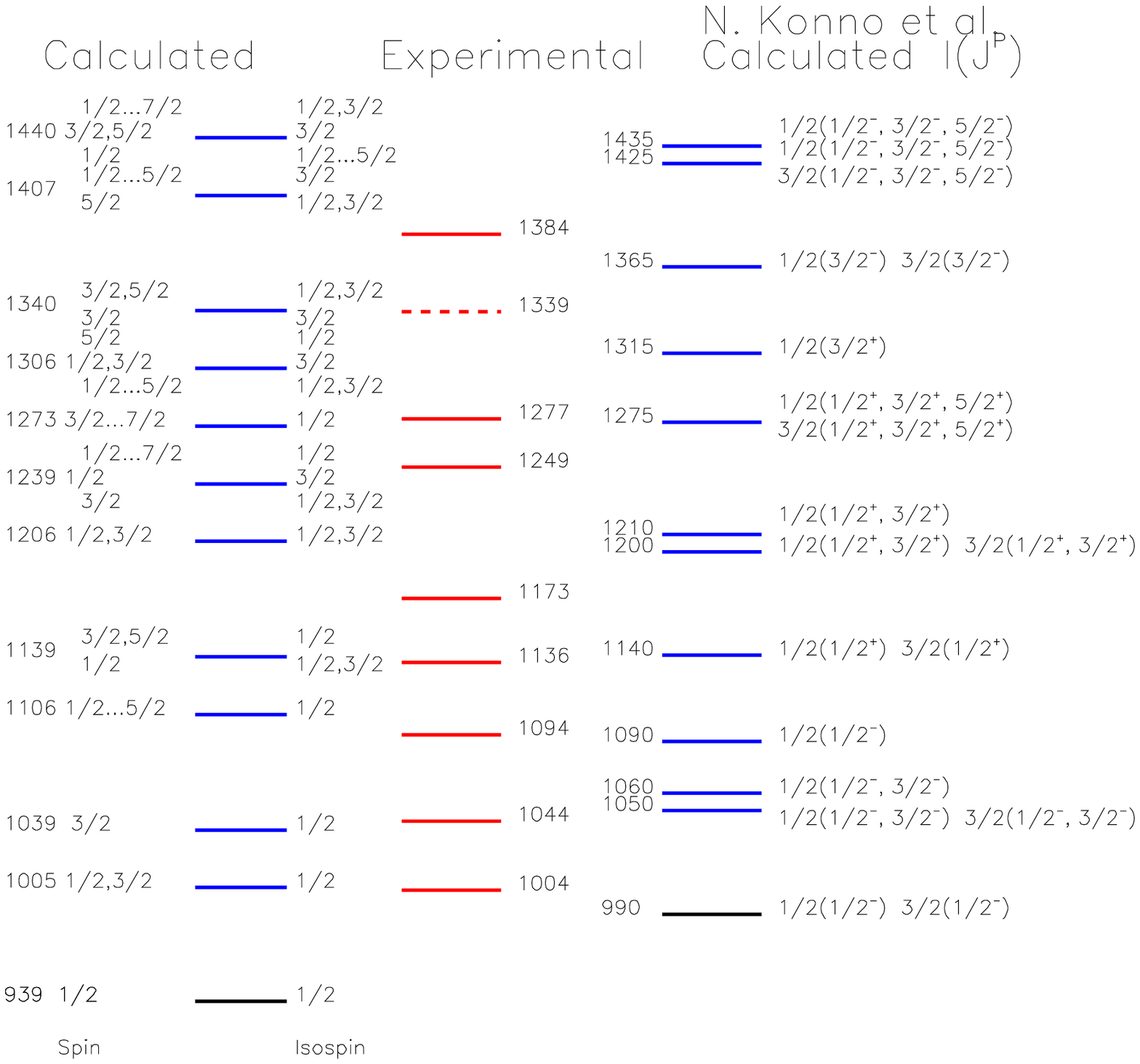}}
\vspace*{-2.mm}
\caption[Fig. 8]{Calculated and experimental narrow exotic baryon level
spectrum in the range 1100~$\le$~M~$\le$~1400 MeV.}
\end{figure}
\end{center}

\begin{center}
\begin{figure}[!hb]
\hspace*{2.mm}

\vspace*{-1.cm}
\scalebox{.9}[.63]{
\includegraphics[bb=35 361 480 770,clip,scale=1.]{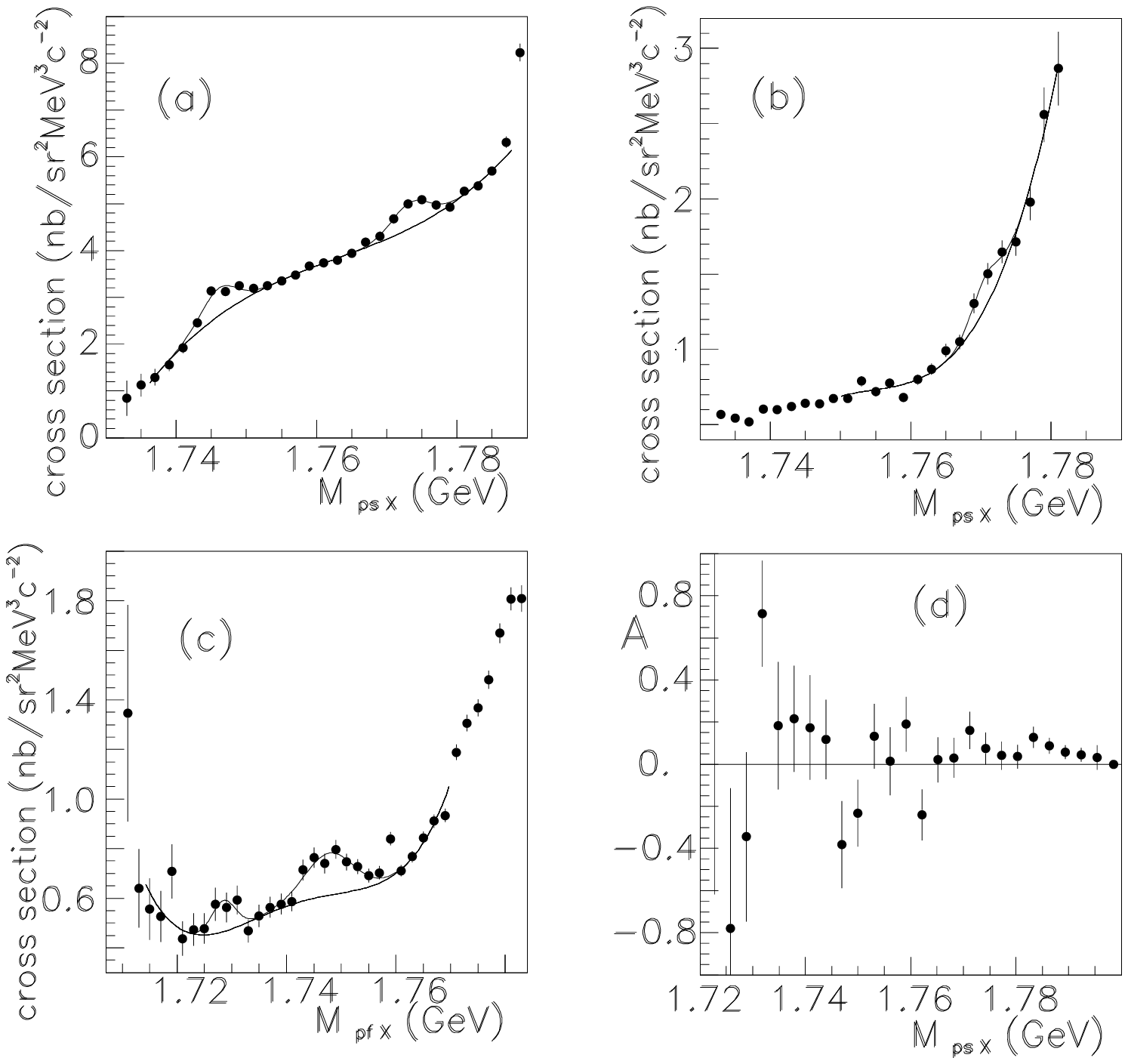}}
\caption[Fig. 9]{Selection of three cross sections and one analyzing power
data showing narrow structures in the mass range 1720$\le$~M~$\le$1800 MeV.}
\end{figure}
\end{center}

\vspace*{-1.7cm}
Several theoretical attempts were performed to explain these
observations:\\
\hspace*{2.mm} A.P. Kobushkin \cite{kob} assumes that the first narrow baryons
(M$\le$1075 MeV) may be members of a total antisymmetric representation of
the spin-flavor group.\\
\hspace*{2.mm} N. Konno, H. Noya, and H. Nakamura \cite{kon} assume that the
exotic baryons consist
of a diquark and a quark (Diquark cluster model).\\
\hspace*{2.mm} Y.E. Pokrovsky \cite{pok} derives a model which deals with
chiral and
scale symmetries of QCD, and describes hadrons as multi-bag states
($Bag \overline{Bag} Bag$) for baryons.\\
\hspace*{2.mm} Th. Walcher \cite{wal} derives a model based on the excitation
of quark
condensate. The model explains the narrow states as a multiple production of
a ``genuine'' Goldstone Boson with a mass close to 20 MeV.\\
\hspace*{2.mm}We tentatively associate our narrow baryons with exotic
baryons made of two coloured quark clusters \cite{bor1} \cite{bor2}.\\

\end{document}